# Approximate Convex Hull of Data Streams

Avrim Blum[*1], Vladimir Braverman[†2], Ananya Kumar[‡3], Harry Lang[§4], and Lin F. Yang[¶5]

1   **TTI-Chicago, Chicago, United States**
    `avrim@ttic.edu`
2   **Johns Hopkins University, Baltimore, United States**
    `vova@cs.jhu.edu`
3   **Carnegie Mellon University, Pittsburgh, United States**
    `skywalker94@gmail.com`
4   **Johns Hopkins University, Baltimore, United States**
    `hlang8@math.jhu.edu`
5   **Princeton University, Princeton, United States**
    `lin.yang@princeton.edu`

## Abstract

Given a finite set of points $P \subseteq \mathbb{R}^d$, we would like to find a small subset $S \subseteq P$ such that the convex hull of $S$ approximately contains $P$. More formally, every point in $P$ is within distance $\epsilon$ from the convex hull of $S$. Such a subset $S$ is called an $\epsilon$-hull. Computing an $\epsilon$-hull is an important problem in computational geometry, machine learning, and approximation algorithms.

In many real world applications, the set $P$ is too large to fit in memory. We consider the streaming model where the algorithm receives the points of $P$ sequentially and strives to use a minimal amount of memory. Existing streaming algorithms for computing an $\epsilon$-hull require $O(\epsilon^{(1-d)/2})$ space, which is optimal for a worst-case input. However, this ignores the structure of the data. The minimal size of an $\epsilon$-hull of $P$, which we denote by OPT, can be much smaller. A natural question is whether a streaming algorithm can compute an $\epsilon$-hull using only $O(\mathsf{OPT})$ space.

We begin with lower bounds that show that it is not possible to have a single-pass streaming algorithm that computes an $\epsilon$-hull with $O(\mathsf{OPT})$ space. We instead propose three relaxations of the problem for which we can compute $\epsilon$-hulls using space near-linear to the optimal size. Our first algorithm for points in $\mathbb{R}^2$ that arrive in random-order uses $O(\log n \cdot \mathsf{OPT})$ space. Our second algorithm for points in $\mathbb{R}^2$ makes $O(\log(\frac{1}{\epsilon}))$ passes before outputting the $\epsilon$-hull and requires $O(\mathsf{OPT})$ space. Our third algorithm for points in $\mathbb{R}^d$ for any fixed dimension $d$ outputs an $\epsilon$-hull for all but $\delta$-fraction of directions and requires $O(\mathsf{OPT} \cdot \log \mathsf{OPT})$ space.

**1998 ACM Subject Classification** F.2.2 Nonnumerical Algorithms and Problems

**Digital Object Identifier** 10.4230/LIPIcs...

---

[*] This work was supported in part by the National Science Foundation under grant CCF-1525971. Work was done while the author was at Carnegie Mellon University.
[†] This material is based upon work supported by NSF Grants IIS-1447639, EAGER CCF-1650041, and CAREER CCF-1652257.
[‡] Now at DeepMind.
[§] This research was supported by the Franco-American Fulbright Commission. The author thanks INRIA (l'Institut national de recherche en informatique et en automatique) for hosting him during the writing of this paper.
[¶] This material is based upon work supported by the NSF Grant IIS-1447639. Work was done while the author was at Johns Hopkins University.

# 1    Introduction

The question addressed by this paper is: *Can we compute approximate convex hulls of data streams using near-optimal space?* Approximate convex hulls are fundamental in computational geometry, computer vision, data mining, and many more (see e.g. [2]), and computing them in a streaming manner is important in the big data regime.

Our notion of approximate convex hulls is the commonly used $\epsilon$-hull. Let $P$ be a set of $n$ points in $\mathbb{R}^d$. Let $C(P)$ denote the convex hull of $P$. We want a small subset $S$ of $P$ such that all points in $P$ are inside $C(S)$ or within distance $\epsilon$ from $C(S)$. Such a set $S$ is also called an $\epsilon$-coreset or $\epsilon$-kernel for $P$ [1, 2]. Since every point in $P$ can be approximated by a sparse convex combination of points in $S$, $S$ is also called a generating set [7].

$\epsilon$-hulls and their variants have been studied extensively in the literature. In the multiplicative error variant, one requires that any directional width (the diameter of $S$ in a particular direction) of $S$ is a $(1 \pm \epsilon)$ approximation to that of $P$. There are subtle differences between these variants, but one can usually change from one variant to another without much effort. For more details, we refer the reader to [2].

Existing work primarily focuses on worst case bounds, which scale poorly with the dimension $d$. The worst case lower bound for the size of an $\epsilon$-hull is $\Omega(\epsilon^{-(d-1)/2})$. Recently, it has been shown in [7] that one can in fact do much better than the worst case bound if the size of the smallest $\epsilon$-hull for $P$ (which we denote as OPT) is small. In their paper, they show that one can efficiently obtain $S$ of size nearly linear in OPT and at most linear in the dimension $d$.

One concern of the algorithms in [7] is that they require storing all points of $P$ in memory. The huge size of real-world datasets limits the applicability of these algorithms. A natural question to ask is whether it is possible to efficiently maintain an $\epsilon$-hull of $P$ when $P$ is presented as a data stream while using a small amount of memory. We provide both negative and positive results, summarized below.

## 1.1    Our Contributions

In Section 3, we show that no streaming algorithm can achieve space bounds comparable to OPT, the optimal size of an $\epsilon$-hull. In particular, under a reasonable streaming model, no streaming algorithm can have space complexity competitive with $f(\text{OPT}, d)$ in 3 dimensions or higher for any $f : \mathbb{N} \times \mathbb{N} \to \mathbb{N}$. This strong lower bound directs us to consider variants on this problem. Note that the lower bound applies to space and not time; for the batch setting, [7] gives a polynomial-time algorithm that computes an $\epsilon$-hull with space $O(d\text{OPT} \log \text{OPT})$.

We devise and prove the correctness of streaming algorithms for three relaxations of the problem. In Section 4, we show the first relaxation, in which the points are from $\mathbb{R}^2$ and come in a random order. In Section 5, we relax the problem (again in $\mathbb{R}^2$) by allowing the algorithm to make multiple passes over the stream. In Section 6, we show the third relaxation, in which the points come in an arbitrary order and from $d$-dimensional space, but we only require to approximate the convex hull in a large fraction of all directions.

In the first relaxation, our algorithm maintains an initially empty point set $S$. When our algorithm sees a new point $p$, it adds $p$ to $S$ if $p$ is at least distance $\epsilon$ away from the convex hull of $S$. Additionally, our algorithm keeps removing points $p' \in S$ when some $p'$ is contained inside the convex hull of $S \setminus \{p'\}$, that is, removing $p'$ does not change the convex hull of $S$. Surprisingly, for any point stream $P$, with high probability this algorithm keeps an $\epsilon$-hull of size $O(\text{OPT} \cdot \log n)$, where $n$ is the size of $P$.

In the second relaxation, we permit the algorithm to make a small number of passes over the stream. Our algorithm begins the first pass by taking $O(1)$ directions and storing the point with maximal dot product with each direction. In each of $O(\log(\frac{1}{\epsilon}))$ subsequent passes, we refine the solution by adding a new direction in sectors that incurred too much error while potentially deleting old directions that become no longer necessary. The algorithm computes an $\epsilon$-hull of size $O(\mathsf{OPT})$.

In the third relaxation, we only need to be correct in "most" directions (all but a $\delta$ fraction of directions). Our algorithm picks $O_d(\frac{\mathsf{OPT}}{\delta^2} \log \frac{\mathsf{OPT}}{\delta})$ random unit vectors. For each of these vectors $v$, we keep the point in the stream that has maximal dot product with $v$. We give a proof based on VC-dimension uniform bounds to show that this algorithm achieves the desired bound.

Although our algorithms are simple, it is surprising that input-dependent bounds are achievable in these settings. To the best of our knowledge, this is the first work that gives streaming algorithms for $\epsilon$-hulls with space complexity comparable to the optimal approximation.

## 1.2 Related Work

**Batch Algorithms for $\epsilon$-kernels**   We use the term batch algorithm for an algorithm that stores the entire set of points in memory. In the batch setting, Bentley, Preparata, and Faust [5] give a $O_d(1/\epsilon^{(d-1)})$ space algorithm for computing an $\epsilon$-hull of a set of points (assuming constant dimension $d$). Agarwal, Har-Peled, and Varadarajan [1] improve the result to give a $O_d(1/\epsilon^{(d-1)/2})$ space algorithm for computing a multiplicative approximation of convex hulls. The running time bounds were further improved in [8, 10, 13]. Recently, Blum, Har-Peled, and Raichel [7] give the only known batch algorithms for an $\epsilon$-hull that are competitive with the optimal $\epsilon$-hull size of the given point set.

**Streaming Algorithms for $\epsilon$-kernels with Worst Case Guarantees**   Hershberger and Suri [11] give a 2-D one-pass streaming algorithm for $\epsilon$-hulls that uses $O(1/\sqrt{\epsilon})$ space. Agarwal, Har-Peled, and Varadarajan [1] give a one-pass streaming algorithm for $\epsilon$-kernels (which is a multiplicative error version of the $\epsilon$-hull) that uses $O_d((1/\epsilon^{\frac{d-1}{2}})\log^d n)$ space. Chan [8] removes the dependency on $n$ and gives a streaming algorithm for $\epsilon$-kernels that uses $O_d((1/\epsilon^{d-3/2})\log^d 1/\epsilon)$ space. This was then improved to $O_d((1/\epsilon^{\frac{d-1}{2}})\log \frac{1}{\epsilon})$ [15] and the time complexity was further improved by Arya and Chan [4]. Chan [9] also gives a dynamic streaming (allowing deletions in the stream) algorithm based on polynomial methods. All of these space bounds assume a constant dimension $d$.

## 2 Preliminaries

▶ **Definition 2.1.** For any bounded set $C \subseteq \mathbb{R}^d$, we say a point $q$ is $\epsilon$-close to $C$ if $\inf_{x \in C} \|q - x\| \leq \epsilon$.

▶ **Definition 2.2.** Given a set of points $P \subseteq \mathbb{R}^n$, $S \subseteq P$ is an $\epsilon$-*hull* of $P$ if for every $p \in P$, $p$ is $\epsilon$-close to the convex hull of $S$.

▶ **Definition 2.3.** Let $\mathsf{OPT}(P, \epsilon)$ denote the size of a (not necessarily unique) smallest $\epsilon$-hull of $P$. We omit $P$ and $\epsilon$ if it is clear from the context.

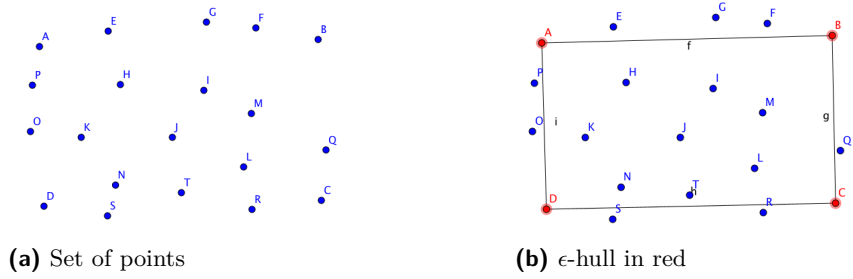

**(a)** Set of points          **(b)** $\epsilon$-hull in red

**Figure 1** $\epsilon$-hull of a set of points.

## 2.1 Streaming Model

Our streaming model, while simple, captures most streaming algorithms for $\epsilon$-hulls in the literature. In our model, a streaming algorithm $\mathcal{A}$ is given $\epsilon \in (0,1)$ in advance but not the size of the input point stream $P \in \mathbb{R}^d$. $P$ is presented to an algorithm $\mathcal{A}$ sequentially:

$$P = (p_1, p_2, \ldots, p_t, \ldots),$$

where $p_t \in \mathbb{R}^d$ is the point coming at time $t$. Note that $P$ may have duplicate points. For the convex hull problem, we require Algorithm $\mathcal{A}$ to maintain a subset $S \subseteq P$. For each point $p \in P$, $\mathcal{A}$ can choose to add $p$ to $S$ (remembering $p$) or ignore $p$ (therefore permanently forgetting $p$). $\mathcal{A}$ can also choose to delete points in $S$, in which case these points are permanently lost. After one-pass of the stream, we require $S$ to be an $\epsilon$-hull of the points set $P$. A trivial streaming algorithm could just keep all points it has seen. However, such an algorithm would not be feasible in the big data regime. Ideally, $\mathcal{A}$ should use space competitive with $\mathsf{OPT}(P, \epsilon)$.

## 3 Lower Bounds

An always-$(f, r)$-optimal algorithm uses space competitive with $f(\mathsf{OPT}(P, \epsilon))$ at all times $t$ and allows the algorithm to maintain an $(r\epsilon)$-hull where $r > 1$. Note that this definition is rather permissive, since it allows an arbitrary function of $\mathsf{OPT}$ and allows slack in $\epsilon$ as well.

▶ **Definition 3.1.** *For $r \geq 1$, $f : \mathbb{N} \times \mathbb{N} \to \mathbb{N}$, we say a streaming algorithm $\mathcal{A}$ is always-$(f, r)$-optimal if given arbitrary $\epsilon > 0$ and point stream $P \subseteq \mathbb{R}^d$, $\mathcal{A}$ keeps an $(r\epsilon)$-hull of $P$ of size at most $f(\mathsf{OPT}(P, \epsilon), d)$.*

▶ **Theorem 3.2.** *For all $r \geq 1$, $d \geq 3$, $f : \mathbb{N} \times \mathbb{N} \to \mathbb{N}$, there does not exist an always-$(f, r)$-optimal streaming algorithm in $\mathbb{R}^d$.*

**Proof.** See Theorem A.5, Theorem A.6, and Corollary A.8 in the Appendix.  ◀

We can also ask a slightly different question: what if an algorithm is given $k$ in advance, and only needs to maintain an $\epsilon$-hull at time $t$ when $\mathsf{OPT}$ of the substream at time $t$ falls below $k$? The algorithm we give for $(\epsilon, \delta)$-hulls in Section 6 is of this form. In the appendix (Definition A.9 and Theorem A.10), we formulate a lower bound for this case.

Our lower bounds guide future research by showing that we need to think beyond the current streaming models, add reasonable assumptions to the problem, or the space bounds of our algorithms must include some functions of $\epsilon$ or $|P|$ (besides just $\mathsf{OPT}$ and $d$).

```
S = {}
When p ∈ P is received do:
  if dist(p, C(S)) ≤ ε:
    // Discard p
  else:
    S = S ∪ {p}
    For each p' ∈ S:
      If p' is an interior point of S then let S = S \ {p'}
```

**Figure 2** Algorithm ROA: pseudocode for 2D random order algorithm.

## 4 2D Random Order Algorithm (ROA)

In many cases, data points are generated i.i.d., for example mixture models or topic models (e.g., [6]). In this section we assume a more general setup: that the points come in a random order. More precisely, for all sets of points $P$, every permutation of $P$ must have equal probability density. The case where the data points are generated i.i.d. (*making no assumptions about the distribution*) is a special case. We assume the points are in 2D. To begin, we introduce the following definition.

▶ **Definition 4.1.** A point $p$ is *interior* to $P$ if $p$ is in the convex hull of $P \setminus \{p\}$.

### 4.1 1D Algorithm

We begin with a classic result in 1-dimension. Consider the algorithm **ROA-insertion**: Begin by keeping a set $S = \{\}$. For each point $p \in P$ that the algorithm sees, if the distance from $p$ to the convex hull of $S$ is at most $\epsilon$, we discard $p$. Otherwise, we add $p$ to $S$.

▶ **Lemma 4.2.** *There exists a constant $c > 0$ such that for any random order input stream $P$ containing at most $n$ points, ROA-insertion maintains a subset $S \subseteq P$ which is an $\epsilon$-hull of $P$ at all times. Moreover, if $P \subseteq \mathbb{R}^1$ then with probability at least $1 - 1/n^3$,*

$$|S| \leq c \cdot 2 \cdot \log n = 2c \cdot \mathsf{OPT}(P, \epsilon) \cdot \log n,$$

*note that for any $\epsilon \geq 0$, $1 \leq \mathsf{OPT}(P, \epsilon) \leq 2$.*

**Proof.** Follows from the classic fact that if $n$ numbers are inserted into a binary search tree in a random order, with high probability the leftmost and rightmost branches have length $O(\log n)$ (e.g., [12]). ◀

A natural question is whether this algorithm generalizes to higher dimensions. Our experiments suggest that the algorithm does not even generalize to 2D. In our experiments, we set $\epsilon = 0$ and gave ROA-insertion $n$ equally spaced points inside a square. OPT is 4, since all the points are contained inside a square. However, experimentally, the number of points kept by ROA-insertion increases much faster than $\log n$.

### 4.2 2D Algorithm

We extend algorithm ROA-Insertion to get algorithm **ROA**. Let the points kept by ROA at the $i^{\text{th}}$ step of the algorithm be $S_i$. At each step $i$, we iteratively delete interior points from $S_i$ until $S_i$ has no interior points. We summarize algorithm ROA in Figure 2.

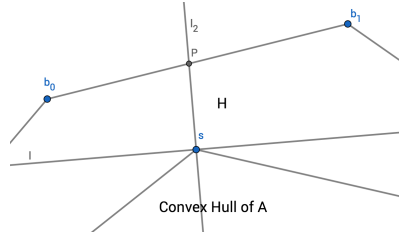

**Figure 3** Figure for Lemma 4.3. Line $l$ separates halfspace $H$ from $\mathcal{C}(A)$. $b_0, b_1$ are on $\partial \mathcal{C}(B)$.

▶ **Lemma 4.3.** *(Similar Boundaries) Suppose $A$ and $B$ are $\epsilon$-hulls of $P$. Then $\mathcal{H}(\partial \mathcal{C}(A), \partial \mathcal{C}(B)) \le \epsilon$, where $\mathcal{H}$ is the (two-way) Hausdorff distance.*

**Proof.** By symmetry, it suffices to show the claim for the one way Hausdorff distance. Consider arbitrary $s \in \partial \mathcal{C}(A)$. We want to show that $\mathrm{dist}(s, \partial \mathcal{C}(B)) \le \epsilon$.

**Case 1** $(s \notin \mathcal{C}(B))$: Then $\mathrm{dist}(s, \partial \mathcal{C}(B)) = \mathrm{dist}(s, \mathcal{C}(B))$. Since $B$ is an $\epsilon$-hull of $P$, $\mathrm{dist}(s, \mathcal{C}(B)) \le \epsilon$.

**Case 2** $(s \in \mathcal{C}(B))$: Refer to Figure 3. Since $s \in \partial \mathcal{C}(A)$, there exists some line $l$ passing through $s$ such that all points in $\mathcal{C}(A)$ lie to one side of $l$. Consider the closed half space $H$ on the other side of $l$. Consider the line $l_2$ perpendicular to $l$, passing through $s$. Since $s \in \mathcal{C}(B)$, $l_2$ intersects $\partial \mathcal{C}(B)$ at some point $p$ that is in $H$. Since $A$ is an $\epsilon$-hull, $\mathrm{dist}(p, \mathcal{C}(A)) \le \epsilon$. However, note that all points in $\mathcal{C}(A)$ are on the other side of line $l$, not containing $H$. So the distance from $p$ to $\mathcal{C}(A)$ is at least the distance from $p$ to $l$, which is the distance from $p$ to $s$. So $\mathrm{dist}(p, s) \le \epsilon$. ◀

▶ **Theorem 4.4.** *There exists a constant $c > 0$ such that for any random order input stream $P$ containing at most $n$ points, ROA maintains a subset $S \subseteq P$ which is an $\epsilon$-hull of $P$ at all times. Moreover, if $P \subseteq \mathbb{R}^2$ then with probability at least $1 - 1/n^2$,*

$$|S| \le c \cdot \mathsf{OPT}(P, \epsilon) \cdot \log n.$$

*Since the algorithm is deterministic, the probability is over the arrival order of $P$.*

**Proof.** An inductive argument shows that at each iteration $i$, $S$ is an $\epsilon$-hull of $P$. We focus on the proof of the space bound.

**Step 1**: We show that all points in $S$ are near the boundary of some optimal $\epsilon$-hull $T$. Note that $S$ does not contain any interior points, so for all $s \in S$, $s \in \partial \mathcal{C}(S)$. Then by Lemma 4.3, for every point $s \in S$, $\mathrm{dist}(s, \partial \mathcal{C}(T)) \le \epsilon$.

**Step 2**: We split the boundary of $T$ into $\mathsf{OPT}$ sections, and show that with high probability our algorithm keeps $O(\log n)$ points for each section. Since $T$ is optimal, it does not contain any interior points. Label the points in $T$: $t_1, ..., t_k$, clockwise along the boundary of the convex hull of $T$. For every $s \in S$, since $\mathrm{dist}(s, \partial \mathcal{C}(T)) \le \epsilon$, $s$ is within distance $\epsilon$ from the line segment connecting some $t_i$ and $t_{i+1}$. Now, referring to Figure X, consider the line segment $l$ connecting arbitrary $t_j$ and $t_{j+1}$, and consider all points within distance $\epsilon$ from $l$. We group the points based on which side of the line segment they are on - consider the points $Q$ on one side of the line segment. The points $Q$ are contained in some narrow strip $R$ with width $\epsilon$. Now, we can apply the proof from Lemma 4.2, by considering the projection of the points in $Q$ onto the line connecting $t_j$ and $t_{j+1}$, to get that with high probability we keep $O(\log n)$ points for each segment.

**Step 3**: We take a union bound over the $\mathsf{OPT}$ sections to get the desired result, where we note that $\mathsf{OPT} \le n$. ◀

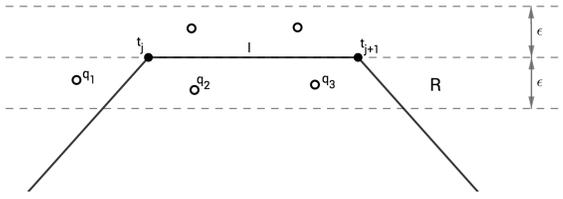

**Figure 4** Figure for Theorem 4.4. Consider all points near segment $l = \overline{t_j t_{j+1}}$. Consider points $q_1, q_2, q_3 \in Q$ on one side of $l$. They are contained in a thin strip $R$ of width $\epsilon$.

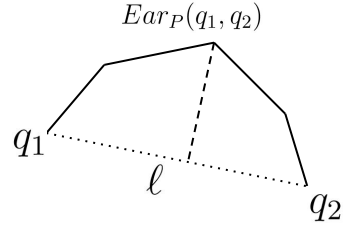

**Figure 5** A diagram of $Ear_P(q_1, q_2)$. The dotted line is $\ell$, and the length of the dashed line is $\mathsf{Error}_P(q_1, q_2)$.

## 5  2D Multipass Algorithm

In this section, we relax the problem by letting the algorithm pass over the stream $P$ multiple times. Let $\mathrm{diam}(P)$ refer to the diameter of the point-set $P$ which is $\max_{x,y \in P} d(x,y)$. Our algorithm requires $\log(\frac{\mathrm{diam}(P)}{\epsilon})$ passes and $O(\mathsf{OPT})$ memory. For convenience, we assume $\mathrm{diam}(P) = 1$ and prove a bound of $\log(\frac{1}{\epsilon})$ passes.

By convention, we define the distance between a point $p$ and a set $A$ to be $d(p, A) = \min_{a \in A} d(p, a)$. We define $\partial P$ to be the subset of $P$ that lies on the boundary of the convex hull of $P$. Formally:

▶ **Definition 5.1.** For a set $P \subset \mathbb{R}^2$, we define $\partial P = P \cap \partial \mathcal{C}(P)$. Here $\partial \mathcal{C}(P)$ means the boundary of the convex hull of $P$.

Given any two points $q_1, q_2 \in \partial P$, define $\ell = \mathcal{C}(\{q_1, q_2\})$ to be the line segment with endpoints $q_1$ and $q_2$. Observe that the set $\mathcal{C}(P) \setminus \ell$ has at most two connected components. Define $Ear_P(q_1, q_2)$ to be the component that lies to the left of the vector from $q_1$ to $q_2$. We define $\mathsf{Error}_P(q_1, q_2) = \max_{x \in Ear_P(q_1, q_2)} d(x, \ell)$ to be the maximum distance of a point in this component from $\ell$. See Figure 5 for an example. Note that we can compute $\mathsf{Error}_P(q_1, q_2)$ in a single pass (see Algorithm 2 and Lemma B.1 in the Appendix).

Let $t$ be a unit vector. Define $\mathsf{GetMax}_P(t)$ to be $\arg\max_{p \in P} p \cdot t$. It is clear that $\mathsf{GetMax}_P(t)$ can be computed in a single pass. Algorithm 1 is the main multipass algorithm, using $\mathsf{Error}$ and $\mathsf{GetMax}$ as blackboxes. We always maintain a set of directions $T$. On Lines 5-9 we run $3|T|$ single-pass algorithms completely in parallel, therefore requiring only a single pass. By the phrase "associating a point with a direction", we mean to keep this point as piece of satellite data.

Our main result for this section is the behavior of Algorithm 1. We define a word as the space required to store a single point in $\mathbb{R}^2$.

We begin with some preliminary statements. We defer the proofs of these lemmas to the Appendix (see Lemmas B.2, B.3, B.4, and B.6). Throughout this section, we use the convention of incrementing subscripts modulo $n$ (for example $q_{n+1} = q_1$).

▶ **Lemma 5.2.** *If Algorithm 1 terminates, it outputs an $\epsilon$-hull to $P$.*

▶ **Lemma 5.3.** *Algorithm 1 terminates in $3 + \lceil \log_2(1/\epsilon) \rceil$ passes.*

▶ **Lemma 5.4.** *Let $p, p', q', q \in \partial P$ be in clockwise order along $\partial \mathcal{C}(P)$. Then $\mathsf{Error}_P(p', q') \leq \mathsf{Error}_P(p, q)$.*

---

**Algorithm 1** Input: a stream of points $P \subset \mathbb{R}^2$ and a value $\epsilon \in (0,1]$. Output: an $\epsilon$-approximate hull of $P$

---

1: $t_1 \leftarrow (1,0)$, $t_2 \leftarrow (-1,0)$
2: $T \leftarrow$ an ordered list $(t_1, t_2)$
3: For $i = \{1, 2\}$, associate $q_i \leftarrow \mathsf{GetMax}(t_i)$ with $t_i$
4: Initialize Flag to down position
5: **for all** $1 \le i \le |T|$ (in parallel) **do**
6: $\quad$ Compute $\mathsf{Error}_P(q_i, q_{i+1})$
7: $\quad$ Compute $\mathsf{Error}_P(q_{i-1}, q_{i+1})$
8: $\quad$ $t'_i \leftarrow$ direction bisecting[1] $t_i$ and $t_{i+1}$
9: $\quad$ $q'_i \leftarrow \mathsf{GetMax}(t'_i)$
10: **for all** $1 \le i \le |T|$ (in parallel) **do**
11: $\quad$ **if** $\mathsf{Error}_P(q_{i-1}, q_{i+1}) \le \epsilon$ and neither $t_{i+1}$ or $t_{i-1}$ have been deleted **then**
12: $\quad\quad$ Remove $t_i$ from $T$
13: $\quad$ **if** $\mathsf{Error}_P(q_i, q_{i+1}) > \epsilon$ **then**
14: $\quad\quad$ Add $t'_i$ to $M$ and associate $q'_i$ with $t'_i$
15: $\quad\quad$ Raise Flag
16: Recompute indices of $T$ to preserve clockwise-order
17: Delete any points/vectors except $t_i \in T$ and their associated $q_i$
18: **if** Flag is up **then**
19: $\quad$ Go to Line 4
20: **else**
21: $\quad$ Output $\{q_1, \ldots, q_{|T|}\}$

---

▶ **Lemma 5.5.** *There exists an $\epsilon$-hull of $P$ using only points from $\partial P$ of cardinality at most $2\mathsf{OPT}(P, \epsilon)$.*

▶ **Theorem 5.6.** *Given a stream of points $P \subset \mathbb{R}^2$ and a value $\epsilon \in (0,1]$, Algorithm 1 terminates within $3 + \lceil \log_2(1/\epsilon) \rceil$ passes, stores at most $24\mathsf{OPT}(P, \epsilon) + O(1)$ words, and returns an $\epsilon$-hull of $P$ of cardinality $6\mathsf{OPT}(P, \epsilon)$.*

**Proof.** Algorithm 1 outputs an $\epsilon$-hull to $P$. By Lemma 5.3, Algorithm 1 terminates after $3 + \lceil \log_2(1/\epsilon) \rceil$ passes. It only remains to bound the space usage and cardinality of the set returned

Let $W \subset \partial P$ be an $\epsilon$-approximation of $P$ such that $n = |W| \le 2\mathsf{OPT}(P, \epsilon)$. Lemma 5.5 guarantees that such a $W$ exists. Let $W = \{w_1, \ldots, w_n\}$ be an ordering of $W$ that is clockwise in $\partial \mathcal{C}(P)$.

By definition, $\epsilon_P(w_i, w_{i+1}) \le \epsilon$ for every $i \in \mathbb{Z}$ (recall the convention of using addition modulo $n$). Consider the state of the algorithm at the beginning of a pass; for notation let $T$ contain the directions $\{t_i\}_{i=1}^{|T|}$ associated respectively with $\{q_i\}_{i=1}^{|T|}$.

For $s \in \{1, 2\}$, suppose that $w_i, q_j, q_{j+s}, w_{i+1}$ are in clockwise order of $\partial \mathcal{C}(P)$. By Lemma 5.4, $\epsilon_P(q_j, q_{j+s}) \le \epsilon_P(w_i, w_{i+1}) \le \epsilon$. We draw two conclusions. The first conclusion ($s = 1$) is that on Line 13, $t'_i$ will not be added to $T$. The second conclusion ($s = 2$) is that on Line 11, $t_{i+1}$ is a candidate for deletion (i.e. $t_{i+1}$ will be deleted unless $t_i$ or $t_{i+2}$ have already been deleted).

Using the clockwise ordering of $\partial \mathcal{C}(P)$, we say that a point $q \in \partial P$ is on edge $(w_i, w_{i+1})$ if it lies between $w_i$ and $w_{i+1}$ in the ordering. Suppose that $\{q_j\}_{j=1}^{|T|}$ contains $m$ points on edge $(w_i, w_{i+1})$. By the reasoning in the preceding paragraph, it is easy to verify that all but $\lceil \frac{m-1}{2} \rceil + 1$ will be deleted on Line 11. As for points added on Line 13, this can only occur at the boundary (between $q_j$ and $q_{j+1}$ where $q_j$ is the last point on some edge) and therefore adds at most 1 point per edge.

Combining these facts, we see that an edge which enters a pass with $m$ points finishes that pass with at most $\lceil \frac{m-1}{2} \rceil + 2$ points. Inductively we begin with $m = \{0, 1, 2\}$ for each edge. This implies that $m \leq 3$ after each pass. Therefore $|T| \leq 3n \leq 6\mathsf{OPT}(P, \epsilon)$ at all times.

Finally, observe that the storage of $4|T| + O(1)$ points are used in a pass. To compute $\epsilon$ without precision issues, storing a single point suffices. Therefore for each $i$ we store one point for each of the two $\epsilon$ computations, one point for GetMax, and the original point $q_i$ and vector $t_i$. The $O(1)$ is just a workspace to carry out the calculations. ◀

## 6  $(\epsilon, \delta)$-Hull

In this section we give an algorithm for a relaxation of $\epsilon$-hulls, which we call $(\epsilon, \delta)$-hulls. Our results hold for arbitrary point sets $P \subseteq \mathbb{R}^d$. Intuitively, an $(\epsilon, \delta)$-hull of $P$ is within distance $\epsilon$ from the boundary of the convex hull of $P$ in at least a $1 - \delta$ fraction of directions.

▶ **Definition 6.1.** Given a vector $v \in \mathbb{R}^d$ and a finite point set $P \subseteq \mathbb{R}^d$, we define the **directional extent** as

$$\omega_v(P) = \max_{p \in P} p \cdot v.$$

If $p \in \mathbb{R}^d$ is a point we define $\omega_v(p) = p \cdot v = \omega_v(\{p\})$. We say that $S$ **maximizes** $P$ in $v$ if $\omega_v(P) = \omega_v(S)$. Note that $S$ can be either a single vector or a set of vectors.

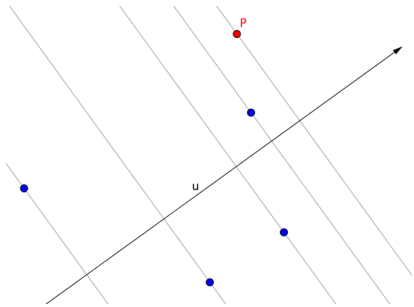

**Figure 6** Point $p$ maximizes the set of points in direction $u$ because its projection onto $u$ is the highest.

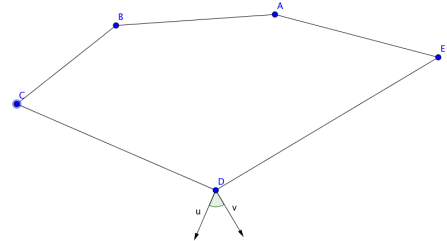

**Figure 7** All vectors between $u$ and $v$ with Euclidean norm at most 1 are in $E_D^T$. See texts for details.

▶ **Definition 6.2.** Given $P \subseteq \mathbb{R}^d$, an $(\epsilon, \delta)$-**hull** is a subset $S \subseteq P$ such that if we pick a vector $v$ uniformly at random from the surface of the unit sphere, $\mathcal{S}^{d-1}$, $S$ $\epsilon$-maximizes $P$ in direction $v$ with probability at least $1 - \delta$, that is,

$$\Pr_{v \sim \mathcal{S}^{d-1}}(|\omega_v(P) - \omega_v(S)| > \epsilon) \leq \delta$$

We start by describing a deterministic 2D algorithm. We pick $O(\frac{k}{\delta})$ uniformly spaced directions, where $k$ is the batch optimal for the $\epsilon$-hull of $P$. For each chosen direction $v$ we store a single point $p$ that maximizes $P$ in direction $v$. It can be verified that these $O(\frac{k}{\delta})$ points give us an $(\epsilon, \delta)$-hull of the input set $P \subseteq \mathbb{R}^2$. Here we focus on higher dimensions. Suppose we fix the dimension $d$. We give a randomized algorithm that uses $m$ points and with probability at least $1 - \gamma$ gives us an $(\epsilon, \delta)$-hull of a point set $P$, where $k$ is the batch optimal for the $\epsilon$-hull of $P$, and $m$ satisfies:

$$m \in O_d\left(\frac{k}{\delta^2} \cdot \log \frac{k}{\gamma \delta}\right),$$

Note that $m$ does not explicitly depend on $\epsilon$. Our algorithm for $d$-dimensional space is as follows: Choose $m$ uniformly random vectors in the unit ball $B^d$ (or equivalently on the surface of the unit ball $\mathcal{S}^{d-1}$). For each chosen vector $v$ we store a single point $p \in P$ that maximizes $P$ in direction $v$, that is, $p \cdot v = \omega_v(P)$. This can easily be done in streaming. Note that the given complexity is for a fixed dimension $d$, the actual space complexity will be multiplied by some (exponential) function of $d$, but independent of $\epsilon$.

## 6.1 Proof of $(\epsilon, \delta)$-hull Algorithm

We begin with some definitions and lemmas.

▶ **Definition 6.3.** Let $B^d$ denote the unit ball in $d$ dimensions. Let $\mathcal{S}^{d-1}$ denote the unit sphere in $d$ dimensions, which is $\partial B^d$ (the boundary of $B^d$).

▶ **Definition 6.4.** Let $V^d(S)$ denote the $d$-dimensional volume (Lebesgue measure) of a measurable set $S$ in $d$-dimensional space.

▶ **Definition 6.5.** Let $P \in \mathbb{R}^d$ be a set of points and $S \subseteq P$. We say $S$ $\epsilon$-**maximizes** $P$ in $v$ if $v = 0$ or, letting $v' = v/|v|_2$, we have

$$|\omega_{v'}(P) - \omega_{v'}(S)| \leq \epsilon.$$

Note that as per definition 6.1, $S$ can be either a single vector or a set of vectors.

▶ **Definition 6.6.** Given $T \subseteq \mathbb{R}^d$ and $t \in \mathbb{R}^d$, we define $E_t^T$ to be the set of all vectors $v \in B^d$ such that $t$ maximizes $T$ in $v$, that is,

$$E_t^T = \{v \mid v \cdot t = \omega_v(T) \wedge |v|_2 \leq 1\}.$$

Figure 7 shows a set of points $T$. All vectors between $u$ and $v$ with Euclidean norm at most 1, in the range indicated by the angle, are in $E_D^T$. Note that $u$ is perpendicular to line segment $CD$ and $v$ is perpendicular to line segment $DE$. Only points $t \in T$ that lie on the boundary of the convex closure of $T$ have non-empty $E_t^T$.

▶ **Definition 6.7.** Given a point stream $P$, and a set $S$, we say *the set of bad vectors* R (with respect to $P, S$) is the set of vectors $v$ in $B^d$ such that $S$ does not $\epsilon$-maximize $P$ in $v$. An equivalent definition of $(\epsilon, \delta)$-hulls is that $V^d(R)/V^d(B^d) \leq \delta$.

We are now ready to present the following lemmas about the properties of $E_t^T$.

▶ **Lemma 6.8** ($\epsilon$-Maximization Lemma). *Suppose $P \subseteq \mathbb{R}^d$ is a finite set of points and $T \subseteq P$ is an $\epsilon$-hull of $P$, and $t \in T$. Then $t$ $\epsilon$-maximizes $P$ for all vectors $v \in E_t^T$ (see Definition 6.6).*

▶ **Lemma 6.9** (Covering Lemma). *For all finite point sets $T \subseteq \mathbb{R}^d$, $\bigcup_{t \in T} E_t^T = B^d$.*

▶ **Lemma 6.10** (Convex Lemma). *For any point $t \in \mathbb{R}^d$ and finite set $T \subseteq \mathbb{R}^d$, $E_t^T$ is convex and has finite volume.*

Next, we provide a crucial lemma that is an analogue to the finite $\epsilon$-net in computational geometry. Our lemma works in continuous measurable space. Before we proceed, For a family of sets $\mathcal{H}$, we denote the simplified version of *VC-dimension* $d' = \widetilde{VC}(\mathcal{H})$ as the smallest positive integer $d'$ such that for every finite set $A \subseteq \mathbb{R}^d$, $|\{h \cap A : h \in \mathcal{H}\}| \leq |A|^{d'}$ (that is, such that Sauer's Lemma holds). We then have the following lemma.

▶ **Lemma 6.11.** *Let $\tau, \gamma \in (0, 1)$ be two parameters. Let $\mathcal{H}$ be a set of measurable sets in $\mathbb{R}^d$ such that $\widetilde{VC}(\mathcal{H}) \leq d'$ for some integer $d'$. Given a measuable convex set $C \subseteq \mathbb{R}^d$, let $\mathcal{H}_C = \{c \in \mathcal{H} : c \subseteq C\}$ be the sets of subsets of $H$ contained in $C$. Suppose we choose $m = \Theta(\frac{d'}{\tau^2} \log \frac{d'}{\tau \gamma})$ points uniformly random in $C$. Then, except with probability $\gamma$, **all** sets $u \in \mathcal{H}_C$ with $V^d(u)/V^d(C) \geq \tau$ contains some selected point, where $V^d(u)$ denotes the volume of $u$.*

**Proof.** Suppose we choose $m$ points $p_1, ..., p_m$ uniformly random in $C$. Consider a set $u \in \mathcal{H}_C$. Let $\mathbb{P}(u)$ denote the probability that a selected point lies in $u$. We have that $\mathbb{P}(u) = V^d(u)/V^d(C)$. Let $\mathbb{P}_m(u)$ denote the empirical estimate of $\mathbb{P}(u)$: $\mathbb{P}_m(u) = \frac{1}{m} \sum_{i=1}^m \mathbb{1}(p_i \in u)$. The VC-dimension of $\mathcal{H}_C$ is bounded by $d'$. By the VC-dimension uniform bound (e.g., [14]),

$$P(\sup_{u \in \mathcal{H}_C} |\mathbb{P}_m(u) - \mathbb{P}(u)| > \tau/2) \leq 8m^{d'} e^{-m\tau^2/128}.$$

Choosing $m = \Theta(\frac{d'}{\tau^2} \log \frac{d'}{\tau \gamma})$ gives us (see Lemma C.1 in the Appendix),

$$P(\sup_{u \in \mathcal{H}_C} |\mathbb{P}_m(u) - \mathbb{P}(u)| > \tau/2) \leq \gamma.$$

So now suppose that $\sup_{u \in \mathcal{H}_C} |\mathbb{P}_m(u) - \mathbb{P}(u)| \leq \tau/2$. Consider arbitrary $u \in \mathcal{H}_C$ with $V^d(u)/V^d(C) \geq \tau$. This means that $\mathbb{P}(u) \geq \tau$. But then $\mathbb{P}_m(u) \geq \tau/2 > 0$. Since $\mathbb{P}_m(u) > 0$, we must have selected at least one point in $u$. By the uniform bound above, except with probability $\gamma$, this is true for **all** $u \in \mathcal{H}_C$. ◀

We want to show that the set of points $S$ our algorithm chooses is $\epsilon$-maximal in most directions. One way is to show that for each point our algorithm picks, the set of *bad vectors* (vectors that our stored points do not $\epsilon$-maximize) shrinks. The next lemma formalizes this notion under some assumptions.

▶ **Lemma 6.12.** *Given a finite point set $P \subseteq \mathbb{R}^d$ and a finite-volume convex set $C \subseteq \mathbb{R}^d$. Assume that there exists some $p \in P$ s.t. for all unit vectors $v \in C$, $p$ $\epsilon$-maximizes $P$ in $v$. Suppose that we pick arbitrary vectors $v_1, ..., v_k \in C$ and corresponding points $p_1, ..., p_k \in P$ s.t. for all $i$, $p_i$ maximizes $P$ in $v_i$. Then there exists a finite-volume convex subset $C' \subseteq C$ s.t.*
1. *For all $i \in [k]$, $v_i \notin C'$.*
2. *For all unit vectors $v \in C \setminus C'$, $S = \{p_1, ..., p_k\}$ $\epsilon$-maximizes $P$ in $v$.*

**Proof.** Consider a vector $v_i$ that we picked, and corresponding point $p_i$. If $p_i = p$, then $C' = \{\}$ satisfies the required conditions. Otherwise, let $H = \{v \mid p_i \cdot v \geq p \cdot v\}$. $H$ is a half-space that contains the vector $v_i$. Furthermore for all vectors $v \in H \cap C$, $S$ $\epsilon$-maximizes $P$ in $v$. So the set of vectors in $C$ that $p_i$ does not maximize are contained in $H^c \cap C$, where $H^c$ does not contain $p_i$. Applying this argument for each vector $v_i$ and corresponding point $p_i$, we can construct $C'$ to be the intersection of $C$ with the $k$ (open) half-spaces corresponding to each of the points $p_i$ we selected. Our constructed $C'$ is convex, because it is the intersection of convex sets, and it is bounded and measurable. ◀

Before we present the main theorem, we show that the set of unions of $k$ ellipsoids is of small VC-dimension. The proof of the following lemma is a simple application of the definition of the VC-dimension with a result bounding VC-dimensions of the set of ellipsoids (e.g. [3]). The formal proof is presented in the Appendix.

▶ **Lemma 6.13.** *Let $\mathcal{E}$ be the sets of all ellipsoids in $\mathbb{R}^d$. Let $\mathcal{E}^k = \{e_1 \cup e_2 \cup e_3 \ldots \cup e_k : e_i \in \mathcal{E}\}$. Then $\widetilde{VC}(\mathcal{H}) \leq 4kd^2$.*

Now we are ready to present the main theorem in this section.

▶ **Theorem 6.14.** *Let $\gamma, \delta \in (0, 1), k \geq 1$ be parameters. Given a point stream $P$ in $\mathbb{R}^d$ and $\epsilon \geq 0$. Suppose $\mathsf{OPT}(P, \epsilon) \leq k$. Then there exists a one-pass streaming algorithm, given $P, \gamma, \delta, k$, stores a set $S \subseteq P$ of $m = \Theta_d\left(\frac{k}{\delta^2} \log \frac{k}{\gamma\delta}\right)$ points, such that, except with probability $\gamma$, $S$ is an $(\epsilon, \delta)$-hull of $P$.*

**Proof.** To begin the proof, we recall the algorithm. We first pick uniformly at random $m = \Theta\left(\frac{d^{2d+2}k}{\delta^2} \log \frac{kd}{\gamma\delta}\right)$ directions from $B^d$, the $d$-dimensional unit ball. When the stream is coming, we maintain the extreme point from $P$ in each direction. The output $S$ is the set of extreme points in each direction.

Intuitively, $S$ is an $(\epsilon, \delta)$-hull iff $B^d$ only contains a small region of bad vectors (with respect to $P$, $S$). Let $T$ be an optimal $\epsilon$-hull of $P$, with $|T| = k$. Fix $t \in T$. Consider the set $E_t^T$. In our proof we will show that with high probability each set $E_t^T$ only contains a small subset of bad vectors, $C'_t$, such that, for all vectors $v \in E_t^T \setminus C'_t$, $S$ $\epsilon$-maximizes $P$ in $v$. Then we show that $\sum_{t \in T} V^d(C'_t) \leq \delta$, which completes the proof.

Suppose the selected random set of vectors is $A \subseteq B^d$. Fix $t \in T$. By Lemma 6.8, $t \in T$ $\epsilon$-maximizes $P$ for all vectors $v \in E_t^T$. Then by Lemma 6.12, there exists a finite-volume convex subset $C'_t \subseteq E_t^T$ such that $C'_t \cap A = \emptyset$ and for all $v \in E_t^T \setminus C'_t$, $S$ $\epsilon$-maximizes $v$. Next, for each $t \in T$, we select a large ellipsoid $u_t$ contained in $C'_t$ such that $V^d(C'_t) \leq V^d(u_t)d^d$. Note that $\cup_{t \in T} C'_t$ is a member of the family $\mathcal{E}^{|T|} = \{h_1 \cup h_2 \cup \ldots \cup h_{|T|} : \forall i, h_i$ is an ellipsoid$\}$. By Lemma 6.13, $\widetilde{VC}(\mathcal{E}^{|T|}) \leq 4kd^2$. By Lemma 6.11, since all $u_t$ do not contain any point from $A$, with probability at least $1 - \gamma$, it must be the case that $V^d(\cup_{t \in T} u_t)/V^d(B^d) \leq \delta/(d^d)$. Since the $u_t$ are disjoint, this means that $V^d(\cup_{t \in T} C'_t)/V^d(B^d) \leq \delta$. Furthermore, by Lemma 6.9, $\bigcup_{t \in T} E_t^T = B^d$. Therefore, with probability at least $1 - \gamma$, $S$ $\epsilon$-maximizes all vectors in $B^d$ except for those in $\cup_{t \in T} C'_t$. Thus, $S$ is an $(\epsilon, \delta)$-hull except with probability $\gamma$.

◀

## 7 Concluding Remarks

In this paper we presented useful lower bounds, and streaming algorithms for relaxations of the $\epsilon$-hull problem that were competitive with $\mathsf{OPT}$. Our work naturally leads to interesting and important open questions. We do not have a lower bound for $\epsilon$-hulls in $\mathbb{R}^2$, so there might exist one-pass streaming algorithm competitive with $\mathsf{OPT}$ in $\mathbb{R}^2$. In $\mathbb{R}^d$, for $d \geq 3$, it might still be possible to devise space-efficient algorithms that include some small function of $\epsilon$ and $\mathsf{OPT}$. Finding an algorithm that scales well with $d$ is an important open problem. Our random order algorithm currently only works in $\mathbb{R}^2$, however our technique could possibly be extended to higher dimensions. Our $(\epsilon, \delta)$-hull algorithm achieves near optimal space bounds for arbitrary constant dimension $d$, where $d$ is small. Note that most of the work on $\epsilon$-kernels also assumes that the dimension $d$ is a constant. We leave it as an open question to remove exponential dependencies on $d$.

# A    Lower Bounds

## A.1   Proof of Always-OPT Lower Bound

▶ **Definition A.1.** *We say a point $p$ is interior to $P$ if $p$ is in the convex hull of $P \setminus \{p\}$.*

▶ **Definition A.2.** *We say a set of points $P$ is meaningful if $P$ has no interior points.*

▶ **Definition A.3.** *For $\epsilon > 0$, we say a set of points $P$ is $\epsilon$-meaningful if the optimal $\epsilon$-hull of $P$ is $P$. This means the distance from point $p \in P$ to the convex hull of $P \setminus \{p\}$ is at least $\epsilon$.*

▶ **Lemma A.4.** *If $P$ is $\epsilon$-meaningful then $P$ is meaningful. In the other direction, if $P$ is meaningful then there exists $\epsilon > 0$ such that $P$ is $\epsilon$-meaningful.*

▶ **Theorem A.5.** *For all $f : \mathbb{N} \times \mathbb{N} \to \mathbb{N}$, there does not exist an always-$(f, 1)$-optimal streaming algorithm in $\mathbb{R}^3$.*

**Proof.** Assume for the sake of contradiction that there exists some function $f : \mathbb{N} \to \mathbb{N}$ and corresponding always-$(f, 1)$-optimal streaming algorithm $A$. Without loss of generality, we can assume that $f$ is increasing.

The high level idea is that we will construct 3 sequences of points $P_1, P_2, P_3$. Let $P_1 \circ P_2$ denote sequence $P_1$ followed by sequence $P_2$ ($P_2$ appended to $P_1$). We will show that if $A$ keeps an $\epsilon$-hull of size at most $f(\mathsf{OPT}(P_1 \circ P_2, \epsilon))$ after receiving $P_1 \circ P_2$, then it cannot keep an $\epsilon$-hull of size at most $f(\mathsf{OPT}(P_1 \circ P_2 \circ P_3, \epsilon))$ after receiving $P_1 \circ P_2 \circ P_3$.

All points in $P_1$ will have $z$-coordinate 0, all points in $P_2$ will have $z$-coordinate $\epsilon$, all points in $P_3$ will have $z$-coordinate $2\epsilon$, where $\epsilon$ will be specified later. Geometrically, one can visualize three planes perpendicular to the $z$ axis with points in $P_1, P_2, P_3$ on their respective planes. We now treat $P_1, P_2, P_3$ as point sets in 2D and specify the $x$ and $y$ coordinates of points in the sets.

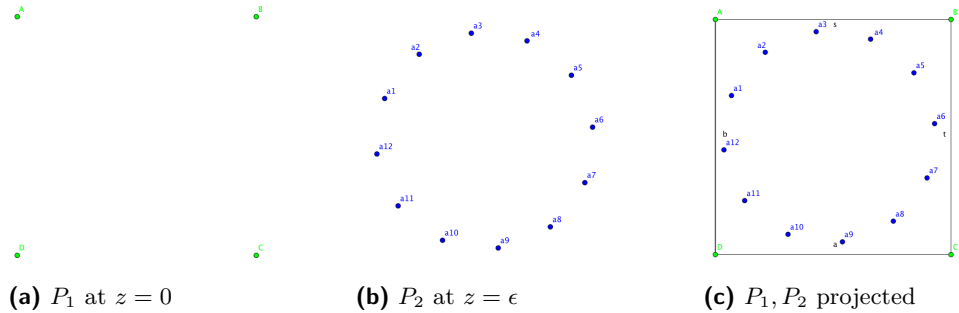

(a) $P_1$ at $z = 0$         (b) $P_2$ at $z = \epsilon$         (c) $P_1, P_2$ projected

**Figure 8** 2D depiction of points in $P_1$ and $P_2$, ignoring $z$ coordinates

$P_1$ contains 4 points that form a square, with coordinates, $(0,0)$, $(0,1)$, $(1,0)$, $(1,1)$, as shown in figure 8a. $P_2$ has $n = 10f(4)$ points, forming a regular polygon that is centered around $(0.5, 0.5)$ with $x, y$ coordinates between 0 and 1, as shown in figure 8b. So if we ignore the $z$ coordinates, $P_2$ is contained inside $P_1$, as shown in figure 8c. Order the points in $P_2$ anti-clockwise, $a_1, ..., a_n$. Group the points into disjoint sets of 5 consecutive points. So the first group will have the points $a_1, ..., a_5$, the second group has the points $a_6, ..., a_{10}$, etc. For each group of 5 points in $P_2$ we will construct $m = 10f(n + 4) = 10f(10f(4) + 4)$ points in $P_3$.

WLOG consider $a_1, ..., a_5$ in $P_2$. For each such group, we will add points $b_1, ..., b_m$ to $P_3$. Ignoring $z$ coordinates, we set $b_1 = a_2$, $b_m = a_4$. All the points $b_1, ..., b_m$ will be contained in the triangle defined by $a_2, a_3, a_4$. The points $b_1, ..., b_m$ are equally spaced, and form equal angles that are less than 180 degrees. We illustrate this construction in figure 9. We note that by our construction $P_1, P_2, P_3$ are meaningful. We choose the smallest $\epsilon$ such that $P_1, P_2, P_3$ are $\epsilon$-meaningful. Now, we will prove the theorem.

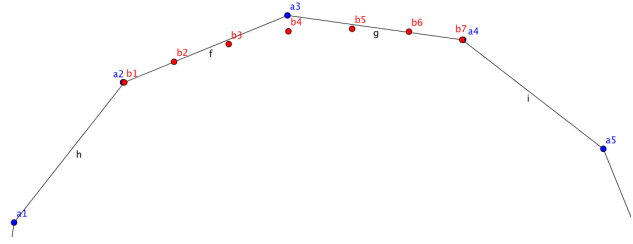

**Figure 9** Red points are in $P_3$, blue in $P_2$

Suppose we run $A$ on $P_1 \circ P_2$. Ignoring $z$ coordinates, $P_2$ is contained inside $P_1$. Since the $z$ coordinates of points in $P_1$ and $P_2$ are 0 and $\epsilon$ respectively, $P_1$ is an $\epsilon$-hull for $P_1 \circ P_2$. So $\mathsf{OPT}(P_1 \circ P_2, \epsilon) \leq 4$. Since $A$ is always-1-$\mathsf{OPT}$, it can keep at most $f(4)$ points in $P_1 \circ P_2$. $P_2$ has a total of $10f(4)$ points, which we divided into $2f(4)$ groups of 5 points. So $A$ did not store any points in at least $f(4)$ of these groups, call these the unselected groups in $P_2$.

Now suppose we run $A$ on $P_1 \circ P_2 \circ P_3$. For each of the $f(4)$ unselected groups in $P_2$, we selected $m = 10f(n+4)$ points in $P_3$. $A$ has to select all the corresponding $m$ points in $P_3$. To prove this, suppose for the sake of contradiction we don't have to select some corresponding point $p$ in $P_3$. $p$ cannot be written as an $\epsilon$-approximate convex combination of selected points in $P_2$, because the $z$ coordinate of $p$ and points in $P_2$ differ by $\epsilon$ *and* if we ignore $z$ coordinates (projecting to the plane $z = 0$) $p$ corresponds to an unselected group and so does not lie inside the convex hull of selected points in $P_2$. Furthermore, since we selected $P_3$ to be $\epsilon$-meaningful, $p$ cannot be written as an $\epsilon$-approximate convex combination of other points in $P_3$.

Summing over unselected groups, this means that $A$ must keep $f(4)m = 10f(4)f(n+4)$ points in $P_3$. The optimal $\epsilon$-hull of $P_1 \circ P_2 \circ P_3$ is much smaller: we can simply store all the points in $P_1$ and $P_2$ giving us a total of $n+4$ points. So $\mathsf{OPT}(P_1 \circ P_2 \circ P_3, \epsilon) \leq n+4$, which means $A$ is allowed to keep at most $f(n+4)$ points. $10f(4)f(n+4) > f(n+4)$ so we have a contradiction.

◂

▶ **Theorem A.6.** *For all $r \in \mathbb{Z}^+$, $f : \mathbb{N} \times \mathbb{N} \to \mathbb{N}$, there does not exist an always-$(f, r)$-optimal streaming algorithm in $\mathbb{R}^3$.*

**Proof.** The construction is similar to theorem A.5, with a few differences. Instead of constructing 3 sets of points $P_1, P_2, P_3$, we construct $r+1$ sets of points $P_1, ..., P_{r+1}$. All points in $P_i$ will have $z$ coordinate $(i-1)\epsilon$. For the $x, y$ coordinates, the construction of $P_{i+1}$ for $i \geq 2$ is similar to the construction of $P_3$ in theorem A.5. We group $P_i$ into disjoint groups of 5 points. Let $n_i$ be the total number of points in $P_1, ..., P_i$. For each group in $P_i$ we add $10f(n_i)$ points in $P_{i+1}$. We then choose $\epsilon$ so that $P_i$ is $r\epsilon$ meaningful for all $i$.

The proof of the construction is also similar to theorem A.5, except we apply the argument inductively. We can show that for each $i$, there exists an unselected group of 5 points in $P_i$, and further if we project onto $z = 0$ the unselected group would be at least distance $r\epsilon$ outside the convex hull of points we select in $P_2, ..., P_{i-1}$. At $P_{r+1}$, this will give us an unselected point that is not an $r\epsilon$-approximate convex combination of selected points. ◂

▶ **Corollary A.7.** *The above proof holds for both deterministic and randomized algorithms since it first presents a construction, and then cases on a particular run of the algorithm. It does not assume the algorithm runs the same way each time.*

▶ **Corollary A.8.** *For all $r \geq 1$, $d \geq 3$, $f : \mathbb{N} \times \mathbb{N} \to \mathbb{N}$, there does not exist an always-$(f, r)$-optimal streaming algorithm in $\mathbb{R}^d$.*

**Proof.** If $d \geq 3$, $\mathbb{R}^d$ contains a 3D subspace so we can use the construction in Theorem A.6.
◀

## A.2 Sometimes-OPT Lower Bound

▶ **Definition A.9.** *Given $f : \mathbb{N} \times \mathbb{N} \to \mathbb{N}$, a streaming algorithm $\mathcal{A}$ is sometimes-$f$-optimal if the following holds. Suppose $\mathcal{A}$ is given $k \in \mathbb{Z}^+$ in advance, and is run on an arbitrary point stream $P$ with $\mathsf{OPT}(P, \epsilon) \leq k$. At all times, $\mathcal{A}$ is allowed to keep at most $f(k)$ points. After processing all points in $P$, $\mathcal{A}$ keeps an $\epsilon$-hull of $P$.*

▶ **Theorem A.10.** *For all $f : \mathbb{N} \times \mathbb{N} \to \mathbb{N}$, $d \geq 3$, there does not exist a sometimes-$f$-optimal streaming algorithm in $\mathbb{R}^d$.*

**Proof.** We first prove this for a deterministic algorithm and then sketch out how to extend the proof to a randomized algorithm. Assume for the sake of contradiction that there exists a deterministic sometimes-OPT streaming algorithm in $\mathbb{R}^3$.

We modify the construction in theorem A.5. We construct 3 points sets $P_1, P_2, P_3$. Points in $P_1, P_2, P_3$ have $z$-coordinates $0, \epsilon, 2\epsilon$ respectively, so we describe their $x, y$ coordinates. $P_1$ contains 4 points $(0,0), (0,1), (1,0), (1,1)$ like in theorem A.5. $P_2$ is a regular polygon centered around $(0.5, 0.5)$ with $x, y$ coordinates between 0 and 1. However, $P_2$ contains $n = 10f(7)$ points. We group the points in $P_2$ into consecutive groups of 5 like in theorem A.5.

We will set $k = 7$ and run $A$ on $P_1 \circ P_2$. $A$ is allowed to keep at most $f(7)$ points, so it can keep at most $f(7)$ points in $P_2$. However, $P_2$ had $2f(7)$ groups of 5 points. So at least one of the groups is unselected, suppose the points in one of these groups are $a_1, ..., a_5$. For this group of 5 points, we use the construction we used in theorem A.5 shown in figure 9, except we add $2f(7)$ (instead of $10f(10f(4) + 4)$) points to $P_3$. If we project all points onto the plane at $z = 0$ then $P_3$ will be contained inside the triangle defined by $a_2, a_3, a_4$.

We define $\epsilon$ to be the smallest value such that $P_1, P_2, P_3$ are $\epsilon$-meaningful. Note that the choice of $\epsilon$ is independent of which group was unselected, since $P_2$ is a regular polygon and is therefore symmetric. Now, suppose we run $A$ on the stream $P_1 \circ P_2 \circ P_3$. $P_1 \cup \{a_2, a_3, a_4\}$ forms an $\epsilon$-hull of $P_1 \circ P_2 \circ P_3$ so $\mathsf{OPT}(P_1 \circ P_2 \circ P_3, \epsilon) \leq 7$. Since $A$ is sometimes-OPT, it must find a way to keep an $\epsilon$-hull of $P_1 \circ P_2 \circ P_3$ of size $\leq f(7)$. However, $A$ must choose all points in $P_3$, because their distance from selected points in $P_1$ and $P_2$ is greater than $\epsilon$, and $P_3$ is $\epsilon$-meaningful. So $A$ must store $2f(7)$ points, a contradiction.

This proof can be extended to show that there does not exist a randomized sometimes-$f$-optimal streaming algorithm in $\mathbb{R}^3$.
◀

## B  Proofs deferred from Multipass Section

Algorithm 2 shows how to compute $\mathsf{Error}_P(q_1, q_2)$ in a single pass. We will use $\mathsf{Error}_P(q_1, q_2)$ to verify whether an $\epsilon$-hull requires another point in between $q_1$ and $q_2$. In the algorithm, note that $Ear_P(q_1, q_2)$ is the set of all points $p \in P$ such that $(q_1, p, q_2)$ is clockwise.

▶ **Lemma B.1.** *Algorithm 2 returns $\mathsf{Error}_P(q_1, q_2)$.*

**Algorithm 2** Input: a stream of points $P \subset \mathbb{R}^2$ and two points $q_1, q_2 \in \partial P$. Output: $\mathsf{Error}_P(q_1, q_2)$

1: $\mathsf{Error} \leftarrow 0$
2: $H \leftarrow \mathcal{C}(\{q_1, q_2\})$
3: **for all** $p \in P$ **do**
4:     **if** $(q_1, p, q_2)$ is clockwise **then**
5:         $\mathsf{Error} \leftarrow \max(\mathsf{Error}, dist(p, H))$
6: Return $\mathsf{Error}$

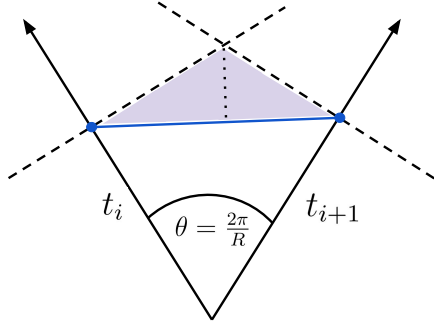

**Figure 10** The two blue points are $\mathsf{GetMax}_P(t_i)$ and $\mathsf{GetMax}_P(t_{i+1})$. The error is at most the height (dotted line) of the shaded triangle. The bass of the triangle (blue line) is at most $diam(P) = 2$. When there are $R$ equally-spaced directions, the angle at the apex of the triangle is exactly $\pi - \frac{2\pi}{R}$. Basic trigonometry shows that height of this triangle (an upper bound on the error) is at most $tan(\frac{\pi}{R})$.

**Proof.** Define $\ell$ to be the line segment with endpoints of $q_1$ and $q_2$. Let $A = Ear_P(q_1, q_2)$. We must show that Algorithm 2 returns $\max_{a \in A} d(a, \ell)$.

Observe that $A$ can be written as $\mathcal{C}(P) \cap X$ where $X$ is the closed half-space containing $A$ such that $\ell \subset \partial A$. Therefore $A$ is also convex. This implies that $\max_{a \in A} d(a, \ell) = \max_{a \in (P \cap A)} d(a, \ell)$ since the maximum must occur at a vertex of $A$ and every vertex of $A$ belongs to $P$. Line 4 filters the stream so that we only consider $P \cap A$, and the result follows. ◀

▶ **Lemma B.2.** *If Algorithm 1 terminates, it outputs an $\epsilon$-hull to $P$.*

**Proof.** Let $q_1, \ldots, q_n \in \partial P$ be the output of Algorithm 1. By Line 13, $\mathsf{Error}_P(q_i, q_{i+1}) \leq \epsilon$ for every $1 \leq i \leq n$. We must show that $\{q_1, \ldots, q_n\}$ is an $\epsilon$-hull to $P$.

Define $H = \mathcal{C}(\{q_1, \ldots, q_n\})$. Consider any $p \in P$. Observe that either $p \in H$ (in which case we are done) or there is some $i \in [n]$ such that $p \in Ear_P(q_i, q_{i+1})$. In the latter case, define $\ell$ to be the line segment with endpoints $q_i$ to $q_{i+1}$. Then $d(p, H) \leq d(p, \ell) \leq \mathsf{Error}_P(q_i, q_{i+1}) \leq \epsilon$. ◀

▶ **Lemma B.3.** *Algorithm 1 terminates in $3 + \lceil \log_2(1/\epsilon) \rceil$ passes.*

**Proof.** Let $\tau_1, \ldots, \tau_R$ be $R$ equally-spaced directions for some $R \geq 4$. Recall that under our (trivially removable) assumption, $diam(P) = 2$. See Figure 10 for a diagram of why

$$\mathsf{Error}(\mathsf{GetMax}(t_i), \mathsf{GetMax}(t_{i+1})) \leq \tan(\pi/R)$$

Observe that after the $y^{\text{th}}$ pass, the set of directions $T$ is a subset of $\{\tau_1, \ldots, \tau_{2^y}\}$. The directions that have been removed are precisely those between directions that have error at most $\epsilon$ (see Line 11). This ensures an error bounded by the above expression when $R = 2^y$. The algorithm terminates when all errors are at most $\epsilon$. Therefore the number of passes is at most $y + 1$ where $y$ is the minimum non-negative integer such that $\tan(\pi/2^y) \leq \epsilon$. Re-arranging shows that

$$y = \left\lceil \log_2 \left( \frac{\pi}{\tan^{-1} \epsilon} \right) \right\rceil$$

and the result follows since $\tan^{-1}(\epsilon) \geq \pi\epsilon/4$ for all $\epsilon \in (0, 1]$. ◄

▶ **Lemma B.4.** *Let $p, p', q', q \in \partial P$ be in clockwise order along $\partial \mathcal{C}(P)$. Then $\mathsf{Error}_P(p', q') \leq \mathsf{Error}_P(p, q)$.*

**Proof.** First observe that $Ear_P(p', q') \subset Ear_P(p, q)$. Let $\ell$ be $\mathcal{C}(\{p, q\})$ and let $\ell'$ be $\mathcal{C}(\{p', q'\})$. It remains to show that for every $x \in Ear_P(p', q')$, $d(x, \ell') \leq d(x, \ell)$.

Let $A$ be the closure of $Ear_P(p, q)$. This is also a convex shape. Now observe that $Ear_A(p', q') = Ear_P(p', q')$. Let $y \in \ell$ be the point such that $d(x, y)$ is minimized. This exists since $\ell$ is compact. $\ell$ and $y$ lie in different connected components of $A \setminus \ell'$. By convexity of $A$, the line segment from $x$ to $y$ must pass through $\ell'$. Therefore $d(x, \ell') \leq d(x, y) = d(x, \ell)$ as desired. ◄

▶ **Lemma B.5.** *Let $P$ be a set in $\mathbb{R}^2$. Then $\mathsf{OPT}(P, \frac{\epsilon}{2}) \leq 6\mathsf{OPT}(P, \epsilon)$.*

**Proof.** Let $S \subset \partial P$ be a set of $2\mathsf{OPT}(P, \epsilon)$ points that forms an $\epsilon$-hull of $P$ (such a set exists by Lemma 5.5). We will construct an $(\epsilon/2)$-hull of $P$ consisting of $3|S|$ points. Let $s_1$ and $s_2$ be two adjacent points after orienting $S$ clockwise. Define $\ell$ to be the line through $s_1$ and $s_2$, and let $H$ be the open half-space with boundary $\ell$ and containing the left-hand side of the directed segment $(s_1, s_2)$ (therefore containing no points of $S$). In the case that $|S| = 1$, we continue the proof by letting $\ell$ be any line such that $\ell \cap \mathcal{C}_P = S$.

Let $Q = H \cap P$, and observe that all points of $Q$ are within distance $\epsilon$ of $\ell$. Every point in $Q$ within distance $\frac{\epsilon}{2}$ of $\ell$ is also within distance $\frac{\epsilon}{2}$ of $\mathcal{C}_{\{s_1, s_2\}}$. Therefore to form an $\frac{\epsilon}{2}$-hull of $P$, we only need to add points to take care of those remaining points in $Q$ that have distance between $\frac{\epsilon}{2}$ and $\epsilon$ to $\ell$, which we denote by $Q'$. Define $\bar{Q}_1$ to be the $\max(2, |Q'|)$ points of $Q'$ that are extremal when projected onto $\ell$. Now every point of $Q$ is within distance $\frac{\epsilon}{2}$ of $\mathcal{C}_{\{s_1, s_2\} \cup \bar{Q}_1}$. Repeat this process on each edge to construct the sets $\bar{Q}_1, \ldots, \bar{Q}_{|S|}$ whose union we denote by $\bar{Q}$. The set $S \cup \bar{Q}$ forms an $\frac{\epsilon}{2}$-hull of size at most $3|S|$. ◄

▶ **Lemma B.6.** *There exists an $\epsilon$-hull of $P$ using only points from $\partial P$ of cardinality at most $2\mathsf{OPT}(P, \epsilon)$.*

**Proof.** Let $Q$ be an $\epsilon$-hull of $P$ such that $|Q| = \mathsf{OPT}(P, \epsilon)$. Choose an arbitrary point $s$ from the interior of $\mathcal{C}(Q)$. For each $q_i \in Q$, define $q'_i$ to be $s + t_i(q_i - s)$ for the unique $t_i > 0$ such that $q'_i \in \partial \mathcal{C}(P)$. Observe that $Q'$ (obtained by replacing each $q_i \in Q$ with $q'_i$) has the property that $\mathcal{C}(Q') \supset \mathcal{C}(Q)$.

Now for each $q'_i \in Q'$, we have that $q'_i \in \partial \mathcal{C}(P)$ and there must be two points $p_i^1, p_i^2 \in \partial P$ such that $q'_i \in \mathcal{C}(\{p_i^1, p_i^2\})$. If $q'_i \in \partial P$ then we simply set $p_i^1 = p_i^2 = q'_i$. Define the set $Q''$ to be the union of the $\{p_i^1, p_i^2\}$ over all the $q'_i$. Since $q'_i \in \mathcal{C}(\{p_i^1, p_i^2\})$, we have that $\mathcal{C}(Q'') \supset \mathcal{C}(Q)$. Therefore $Q''$ is an $\epsilon$-hull of $P$, $|Q''| \leq 2|Q|$, and $Q'' \subset \partial P$. ◄

## C Proofs Deferred from $(\epsilon, \delta)$-hull Section

▶ **Lemma C.1.** *If $d' \geq 1$, $r \leq 1$, $0 < \gamma \leq 1$, and $m = 3 \cdot \frac{32}{r^2} \cdot d' \cdot \log(\frac{32}{r^2} \cdot d' \cdot \frac{1}{\gamma})$, then $8m^{d'} e^{-mr^2/32} \leq \gamma$.*

**Proof.** Let $l = \log(\frac{32}{r^2} \cdot d' \cdot \frac{1}{\gamma})$.

$$
\begin{aligned}
8m^{d'} e^{-mr^2/32} &= 8(3 \cdot \frac{32}{r^2} \cdot d' \cdot l)^{d'} e^{-mr^2/32} && \text{[Substituting } m\text{]} \\
&= 8(3 \cdot \frac{32}{r^2} \cdot d' \cdot l)^{d'} e^{-3 \cdot d' \cdot l} && \text{[Substituting } m\text{]} \\
&= 8(3 \cdot \frac{32}{r^2} \cdot d' \cdot l)^{d'} ((e^{-l})^{d'})^3 && \text{[Manipulating exponent]} \\
&= 8(3 \cdot \frac{32}{r^2} \cdot d' \cdot l)^{d'} ((\frac{r^2}{32} \cdot \frac{1}{d'} \cdot p)^{d'})^3 && \text{[Substituting } l\text{]}
\end{aligned}
$$

We split the left half of the expression into three parts:

$$8(3 \cdot \frac{32}{r^2} \cdot d' \cdot l)^{d'} = [8 \cdot 3^{d'}] \cdot [(\frac{32}{r^2} \cdot d')^{d'}] \cdot [l^{d'}]$$

We separately multiply each part by $(\frac{r^2}{32} \cdot \frac{1}{d'} \cdot \gamma)^{d'}$ to get the desired result.

$$
\begin{aligned}
[8 \cdot 3^{d'}] \cdot (\frac{r^2}{32} \cdot \frac{1}{d'} \cdot \gamma)^{d'} &\leq (24)^{d'} \cdot (\frac{r^2}{32} \cdot \frac{1}{d'} \cdot \gamma)^{d'} && [1 \leq d' \text{ so } 8 \leq 8^{d'}] \\
&\leq (24)^{d'} \cdot (\frac{1}{32} \cdot 1 \cdot 1)^{d'} && [\text{Since } r \leq 1, 1 \leq d', \text{ and } \gamma \leq 1] \\
&\leq 1
\end{aligned}
$$

$$
\begin{aligned}
[(\frac{32}{r^2} \cdot d')^{d'}] \cdot (\frac{r^2}{32} \cdot \frac{1}{d'} \cdot \gamma)^{d'} &= \gamma^{d'} \\
&\leq \gamma
\end{aligned}
$$

$$
\begin{aligned}
[l^{d'}] \cdot (\frac{r^2}{32} \cdot \frac{1}{d'} \cdot \gamma)^{d'} &= [(\log(\frac{32}{r^2} \cdot d' \cdot \frac{1}{\gamma}))^{d'}] \cdot (\frac{r^2}{32} \cdot \frac{1}{d'} \cdot \gamma)^{d'} \\
&\leq [(\frac{32}{r^2} \cdot d' \cdot \frac{1}{\gamma})^{d'}] \cdot (\frac{r^2}{32} \cdot \frac{1}{d'} \cdot \gamma)^{d'} && [\text{Since } \log x \leq x] \\
&= 1
\end{aligned}
$$

◀

**Proof of Lemma 6.13.** We consider an arbitrary finite set $A \subset \mathbb{R}^d$. Denote $\mathcal{E}_A = \{e \cap A : e \in \mathcal{E}\}$ and $\mathcal{E}_A^k = \{e \cap A : e \in \mathcal{E}^k\}$. Consider an $e = e_1 \cup e_2 \cup e_3 \ldots \cup e_k \in \mathcal{E}^k$. We have

$$A \cap e = \cup_{j=1}^{k} (e_j \cap A).$$

Since each $e_j \cap A \in \mathcal{E}_A$, we thus have

$$|\mathcal{E}_A^k| \leq |\mathcal{E}_A|^k.$$

By [3], we have that $\text{VC}(\mathcal{E}) \leq 2d^2$ and $|\mathcal{E}_A| \leq (|A|+1)^{2d^2} \leq |A|^{4d^2}$. Thus we have

$$|\mathcal{E}_A^k| \leq |A|^{4kd^2},$$

which completes the proof. ◀